\documentclass[showpacs,superscriptaddress,10pt,twocolumn]{revtex4-2} 
\usepackage{amsmath,amssymb,graphicx}
\usepackage{blindtext}
\usepackage{epstopdf}
\usepackage{xcolor}
\begin{document}
\setlength{\parskip}{0pt}
\title{Deterministic preparation of $W$ States via spin-photon interactions}

\author{Fatih Ozaydin} 
\affiliation{Institute for International Strategy, Tokyo International University, 1-13-1 Matoba-kita, Kawagoe, Saitama, 350-1197, Japan}

\author{Can Yesilyurt}\email{Corresponding author: can--yesilyurt@hotmail.com}
\affiliation{Department of Physics, Faculty of Science, Istanbul University, 34116, Vezneciler, Istanbul, Turkey}

\author{Sinan Bugu}
\affiliation{Department of Electrical and Electronic Engineering, Tokyo Institute of Technology, 2-12-1 Ookayama,
Meguro-ku, Tokyo 152-8552, Japan}

\author{Masato Koashi}
\affiliation{Photon Science Center, Graduate School of Engineering, The University of Tokyo, 7-3-1 Bunkyo-ku, Tokyo 113-8656, Japan}


\begin{abstract}
Spin systems such as silicon or nitrogen vacancy centers in diamond, quantum dots and quantum dot molecules coupled to optical cavities appear as key elements for creating quantum networks
as not only constituting the nodes of the network, but also assisting the creation of photonic networks. Here we study deterministic preparation of arbitrary size $W$ states with spin systems.
We present an efficient operation on three qubits, two being the logical qubits and one being the ancillary qubit, where no interaction between the logical qubits are required.
The proposed operation can create a $W$-type Einstein-Podolsky-Rosen (EPR) pair from two separable qubits, and expand that EPR pair or an arbitrary size $W$ state by one, creating a $W$-like state.
Taking this operation as the fundamental building block, we show how to create a large scale $W$ state out of separable qubits, or double the size of a $W$ state.
Based on this operation and focusing on nitrogen vacancy (NV) centers in diamond as an exemplary spin system, we propose a setup for preparing $W$ states of circularly polarized photons, assisted by a single spin qubit, where no photon-photon interactions are required.
Next, we propose a setup for preparing $W$ states of spin qubits of spatially separated systems, assisted by a single photon. 
We also analyze the effects of possible imperfections in implementing the gates on the fidelity of the generated $W$ states.
In our setups, neither post-measurement, nor post-processing on the states of spin or photonic qubit is required.
Our setups can be implemented with current technology, and we anticipate that they contribute to quantum science and technologies.

\end{abstract}


\maketitle

\section{Introduction}
Preparing multipartite entangled systems belonging to basic classes such as GHZ~\cite{Greenberger1989}, cluster~\cite{Briegel2001}, Dicke~\cite{Dicke1954} and  $W$~\cite{Dur2001}, is not only a critical step for enabling quantum technologies but also vital for understanding the quantum entanglement from a fundamental perspective.
The preparation methods for cluster and GHZ states are usually straightforward~\cite{Briegel2001,NC} with ongoing efforts on various physical systems such as quantum dot spins~\cite{Ming18LPL}.
On the other hand, an intense theoretical and experimental effort has been devoted to develop efficient methods for preparing $W$ states~\cite{Kobayashi2014,Tashima2008,Tashima2009A,Tashima2009B,Ikuta2011,Tashima2010,Augusiak18PRA,ASharma20PRA,Wei2020NJP}.

This is because an $n$-qubit cluster state in a generic form can be prepared via a series of controlled gates and single qubit rotation gates~\cite{Briegel2003PRA,Tame09PRA}, and a GHZ state in the form
\begin{equation}
	|\text{GHZ}\rangle = \frac{ |0\rangle^{\otimes n} + |1\rangle^{\otimes n} } {\sqrt{2}}
\end{equation}
can be prepared via a series of controlled-NOT gates and Hadamard gates ~\cite{NC}, efficiently.
However, efficient preparation of an $n$-qubit $W$ state requires more efforts due to its sophisticated form~\cite{Ozdemir2011}

\begin{equation}
	|W_n\rangle = {1 \over \sqrt{n} } ( |0\rangle^{\otimes (n-1)} |1\rangle + \sqrt{n-1} | W_{n-1} \rangle |0 \rangle),
\end{equation}

\noindent which reads for $n=4$ as follows
\begin{equation}
	|W_4\rangle = {1 \over \sqrt{4} } ( |0001\rangle + |0010\rangle + |0100\rangle + |1000\rangle).
\end{equation}

A probabilistic fusion setup for polarization based encoded photons was proposed~\cite{Ozdemir2011} and further improved for increasing the probability of success and simultaneous fusion of several $W$ states~\cite{Bugu2013A,Yesilyurt2013A,Ozaydin2014A,Ming16Teleportation,Diker15,Ming16PRA,Ming17QIP,Li18QIP}.
In a recent work, a quantum eraser model was proposed to enhanced the $W$ state fusion process~\cite{Kim2020PRA}.
 
Fusion approach was applied to cavity-QED systems for creating $W$ states of atoms~\cite{CavityRefs0,CavityRefs1,CavityRefs2,CavityRefs21,RecentW4} and quantum dot spins~\cite{RecentW1,RecentW2}. Recently, it was shown that a double quantum dot system can be used for fusing two $W$ states of spin qubits using Pauli spin blockade, requiring no cavities and ancillary photons~\cite{Bugu2020SREPPSB}.

Deterministic expansion schemes were proposed to prepare $W$ states of 4-qubits~\cite{DetExp4} and finally large-scale $W$ states of arbitrary number of polarization based encoded photonic qubits~\cite{DetExp}. This strategy was then considered in preparing atomic qubits via cavities~\cite{CavityRefs21}, and in implementing efficient algorithms on IBM quantum computer~\cite{DetExpR1}.

On the other hand, spin systems are of great importance in quantum science and technologies. 
However, a deterministic expansion strategy for spin systems based on spin-photon interactions is missing. 
Addressing this problem in the present paper, we propose an optimum three-qubit operation and present its circuit model decomposed into only two- and single-qubit gates.
The operation applies on two logical qubits and one ancillary qubit, such that through the ancillary qubit, the desired interaction between the logical qubits is realized without any direct interaction between them.

Then, we first show how this three-qubit operation can be implemented for a system where an ancillary spin qubit assists preparing a $W$ state of circularly polarized photons.
Next, we present the essential contribution of this work, i.e., how to create large-scale $W$ states of spatially separated spins qubits, using an ancillary photon.

The proposed scheme can be realized in various spin systems that possess spin selective reflectivity. Quantum-dot-based technologies and NV centers are the most common platforms in which our proposed method can be applied. The required tasks to prepare $W$ states can be achieved by sequential detection of auxiliary photons in a system with quantum dots or quantum dot molecules coupled to plasmon or optical cavities~\cite{Weiss2019,Cui2019}. Similarly, instead of quantum dots, spin states in solid-state materials such as silicon-vacancy centers in diamond ~\cite{Sukachev2017,Zhang2020} and silicon carbide~\cite{Widmann2019,Miao2019} can be considered as alternative platforms. Besides, various novel phenomena such as spin selective metasurfaces~\cite{Huang2021}, chiral molecules~\cite{Naaman2020}, chiral metamirrors~\cite{Ling2017}, and chiral coupling of valley excitons~\cite{Chen2020} can be utilized to perform the operations required in the present scheme.
Achieving high coherence times even in the room temperature, NV centers coupled to high quality optical cavities enable significant advances in quantum information science.
Including the advantages and excluding the disadvantages of optical and solid state systems, NV centers are considered as strong candidates for distributed computing and quantum communication among large-scale quantum networks~\cite{Wrachtrup04PRL,Lukin06Science,Wrachtrup08Science,Smith09LPR,Hanson13MRS}.
Based on the interaction between an incident photon and the NV center, it has been shown that it is possible to create entanglement and even realize basic two- and three-qubit gates between the electronic spins of spatially separated NV centers~\cite{Lukin10Nature,Du11PRA,Twamley12OptExp,Deng13PRA,Cheng2013JOSAB,Wu13OptExp,Yeon13OptExp,Zhang13JOSAB,Long15PRA,Wang15JOSAB,Deng15PRA,Long15SRep}.
Very recently, the idea of fusing two existing $W$ states to create a larger one has been studied for NV centers, opening a new direction in this field~\cite{Wang17OptExp}.
Hence, we choose NV centers to give an exemplary implementation of our general scheme based on spin-photon interactions in microcavities.

This paper is organized as follows:
In Section II, we present the three-qubit operation in consideration.
In Section III, we present the physics of the NV center briefly as the exemplary spin system, and the interaction between an incident photon and the spin which realizes a controlled-Z gate.
In Section IV and V, we present the implementations of this operation to create $W$ states of circularly polarized photons and spins qubits, respectively.
In Section VI, we analyze how possible imperfections in implementing the single- and two-qubit gates affect the fidelity of the generated $W$ state.
In Section VII, we discuss some advantages and drawbacks of the present scheme and conclude.

\section{A three-qubit operation for deterministic preparation of $W$ states}
\begin{figure}[t]
	\centerline{\includegraphics[width=0.65\columnwidth]{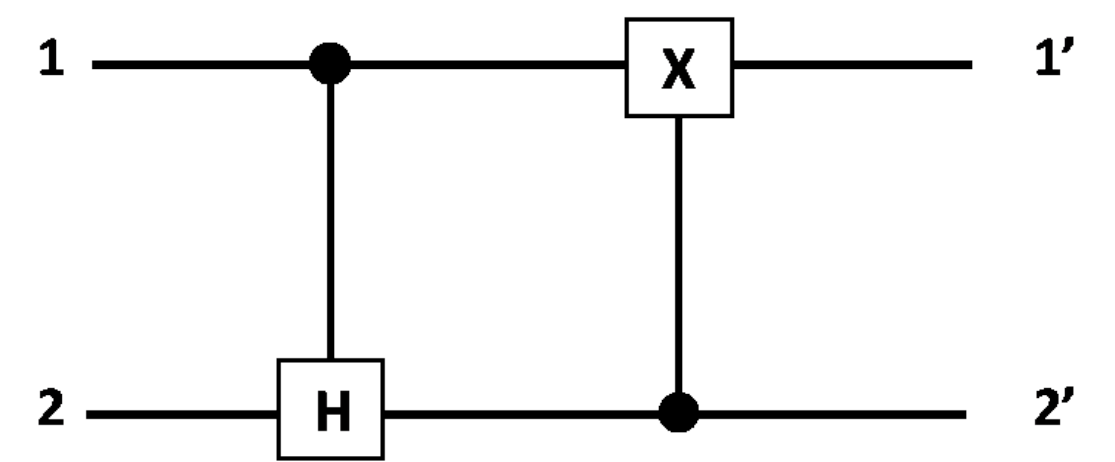}}
	\caption{$W$ state expansion circuit consisting of a controlled-Hadamard gate followed by a controlled-NOT gate, as introduced in Ref.\cite{DetExp}. With the i'th qubit of a $|W_n\rangle$ state in input 1 (for $i=1, 2, ..,n$), a qubit in $|0\rangle$ state in input 2 joins to the $W$ state, creating an $n+1$ qubit $W$-like state. Repeating the procedure for each qubit of the initial $|W_n\rangle$ state, a $|W_{2n}\rangle$ state is created.}\label{fig:JOSABDetExp}
\end{figure}

In the common scenario~\cite{Lukin10Nature,Du11PRA,Deng13PRA,Cheng2013JOSAB,Wu13OptExp,Yeon13OptExp,Zhang13JOSAB,Long15PRA}, a three-qubit operation can be considered for an ancillary photon and two spatially separated spin qubits to be entangled.
We will introduce such a three-qubit operation as the basic building block, show how it can be realized and used for preparing arbitrary size $W$ states of spins qubits.
We will also show that this operation can be used for preparing photonic $W$ states, such that instead of photon-photon interactions, each photon interacts only with the spin qubits separately.
This operation can be considered as an optimization of three-qubit-extension of the two-qubit deterministic expansion circuit presented in Ref.~\cite{DetExp} which consists of a controlled-Hadamard and a controlled-NOT gate as shown in Fig.~\ref{fig:JOSABDetExp}.
With a qubit in state $|1\rangle$ in input 1 and a qubit in state $|0\rangle$ in input 2, this circuit performs the transformation $|10\rangle \rightarrow  (|01\rangle+|10\rangle)/ \sqrt{2}$.
However, as will be detailed later in this section, if the qubit in input 1 belongs to a $|W_n\rangle$ state, the transformation results in an $n+1$ qubit $W$-like state.
Repeating the procedure for each qubit of the initial $|W_n\rangle$ state (with an additional qubit in $|0\rangle$ state in input 2), a genuine $|W_{2n}\rangle$ state is prepared.

Extending a general two-qubit operation -consisting of two controlled gates- to a three qubit operation where the logical qubits do not interact directly but the interaction is realized through an ancillary qubit, requires two swap operations.
Each swap operation can be realized by three two-qubit gates (between the ancillary qubit and the logical qubits), summing up to eight two-qubit gates.
However, because the input qubits of the deterministic expansion operation in consideration have no general but some specific states, that is, not
$\{|00\rangle,|01\rangle,|10\rangle,|11\rangle \}$ but $\{|00\rangle,|10\rangle \}$, it is possible to remove half of the two-qubit gates to realize this operation.

\begin{figure}[t]
	\centerline{\includegraphics[width=1\columnwidth]{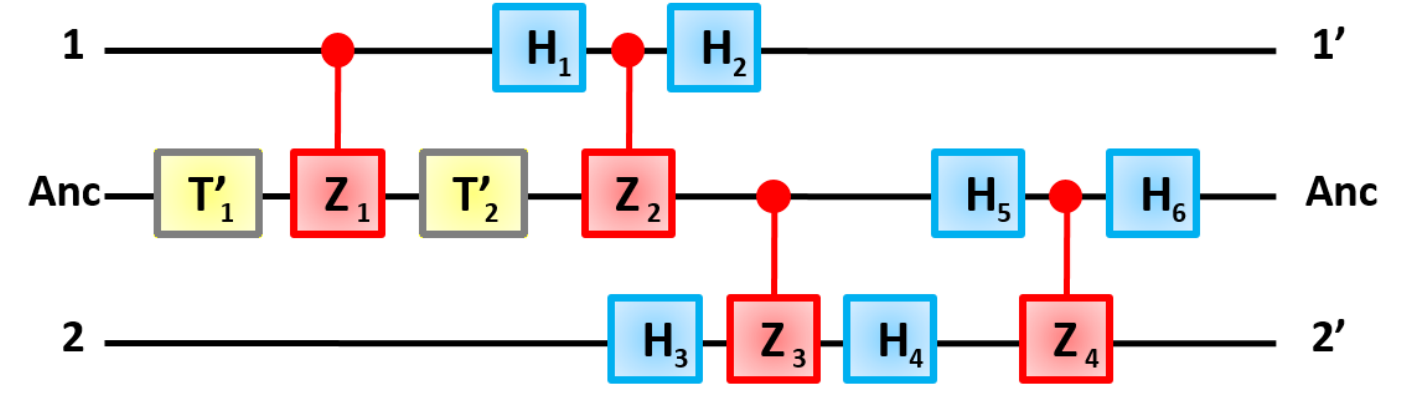}}
	\caption{Three-qubit circuit consisting of only two- and single- qubit gates for implementing the basic expansion operation $O$ in Eq.~\ref{eq:operation}. A qubit of the $W$ state to be expanded is sent through input 1, a qubit in state $|0\rangle$ in the computational basis to join the $W$ state through input 2, and the ancillary qubit in state $|0\rangle$ is at $Anc$ input. $H$ represents Hadamard gate, and $T'$ gate is defined in the text. Controlled-Z gates apply the operation $|1\rangle |1\rangle \rightarrow - |1\rangle |1\rangle$. Subscripts denote the order of applying each gate of the same type.}\label{fig:NVabst}
\end{figure}

As illustrated Fig.~\ref{fig:NVabst}, we consider two logical qubits in the inputs $1$ and $2$, and an ancillary qubit in input $Anc$.
The circuit for the operation consists of four controlled-Z (CZ) gates and eight single qubit gates, i.e. six Hadamard gates and two $\text{T}'$ gates with the operation

\begin{equation}
\text{T}' = \left(
      \begin{array}{cc}
        \cos\theta & \sin\theta \\
        \sin\theta & -\cos\theta \\
      \end{array}
    \right),
\end{equation}
\noindent for $\theta={ \pi \over 8}$.
It is straightforward to show that the overall operation on three qubits in the computational basis is

\begin{equation}
\text{O} =
\left(
  \begin{array}{cccccccc}\label{eq:operation}
    1 & 0 & 0 & 0 &     0 & 0              & 0 & 0 \\
    0 & 0 & 0 & 0 &     {1 \over \sqrt{2}} & 0 &     -{1 \over \sqrt{2}} & 0 \\
    0 & 0 & 0 & 0 &     0 & {1 \over \sqrt{2}} &     0 & -{1 \over \sqrt{2}} \\
    0 & 1 & 0 & 0 &     0 & 0 &     0 & 0 \\
    0 & 0 & 0 & 0 &     {1 \over \sqrt{2}} & 0 &     {1 \over \sqrt{2}} & 0 \\
    0 & 0 & 1 & 0 &     0 & 0 &     0 & 0 \\
    0 & 0 & 0 & 1 &     0 & 0 &     0 & 0 \\
    0 & 0 & 0 & 0 &     0 & {1 \over \sqrt{2}} &     0 & {1 \over \sqrt{2}} \\
  \end{array}
\right).
\end{equation}

\noindent Applying this operation and tracing out the ancillary qubit, a $W$-type Bell pair in the form ${1 \over \sqrt{2}}(|01\rangle+|10\rangle)$ can be created from a three qubit system initially in the separable state $|1\rangle|0\rangle|0\rangle$ in inputs $1$, $Anc$ and $2$, respectively. 
In more detail, the transformation for the initial state $|1\rangle |0\rangle |0\rangle$ is described as 

\begin{eqnarray}\label{eq:3a}
&&\hskip-0.8cm \text{T}'_1, \text{CZ}_1, \text{T}'_2  \to |1\rangle \left({ |0\rangle + |1\rangle \over \sqrt{2} }\right) |0\rangle,  \nonumber \\
&&\hskip-0.8cm \text{H}_1, \text{CZ}_2, \text{H}_2 \to \left({  |1\rangle|0\rangle + |0\rangle|1\rangle  \over \sqrt{2} }\right)  \left(|0\rangle\right),  \nonumber \\
&&\hskip-0.8cm \text{H}_3, \text{CZ}_3, \text{H}_4 \to \left({  |1\rangle|0\rangle|0\rangle + |0\rangle|1\rangle|1\rangle \over \sqrt{2} }\right), \\
&&\hskip-0.8cm \text{H}_5, \text{CZ}_4, \text{H}_6 \to \left({  |1\rangle|0\rangle|0\rangle + |0\rangle|0\rangle|1\rangle \over \sqrt{2} }\right), \nonumber \\
&&\hskip+0.1cm \text{tr}_{\text{Anc}} \to \left({  |1\rangle|0\rangle + |0\rangle|1\rangle \over \sqrt{2} }\right),  \nonumber
\end{eqnarray}


\noindent that is, although the two qubits in input 1 and input 2 have never interacted directly, a two-qubit $W$ type Bell pair is created between them.
Now, let us consider that the qubit in input 1 
is the i'th qubit of an $n$-qubit $W$ state in the form 
$|W_n\rangle ={1 \over \sqrt{n} } \sum_{i=1}^n \bigotimes_{j =1 }^n | \delta_{i j}\rangle$.
Then, this operation expands the state with the qubit in input 2 (denoted as $|0\rangle_2$), creating an $n+1$ qubit $W$-like state as follows.
The i'th term of the initial state with superposition coefficient $1/\sqrt{n}$ splits into two terms with superposition coefficients each $1/\sqrt{2n}$, as 

\begin{eqnarray}\label{eq:term_transformation}
      { | 0\rangle^{\otimes i-1} 		 \otimes |1_i\rangle \otimes |0\rangle_2 		\otimes | 0\rangle^{\otimes n-i} \over \sqrt{n}} \nonumber \\
\to { | 0\rangle^{\otimes i-1}  	  	\otimes |1_i\rangle \otimes |0\rangle_2		\otimes | 0\rangle^{\otimes n-i}  \over \sqrt{2n}} \\
+   { | 0\rangle^{\otimes i-1}  	  	\otimes |0_i\rangle \otimes |1\rangle_2		\otimes | 0\rangle^{\otimes n-i} \over \sqrt{2n}}. \nonumber
\end{eqnarray}



Taking this expansion operation as the basic building block, it is easy to see that applying it to each qubit of the initial $n$-qubit $W$ state with a new qubit in input 2, a $2n$-qubit $W$ state is created, having $2n$ terms all with coefficients $1/\sqrt{2n}$.

\begin{figure}[t]
	\centerline{\includegraphics[width=0.95\columnwidth]{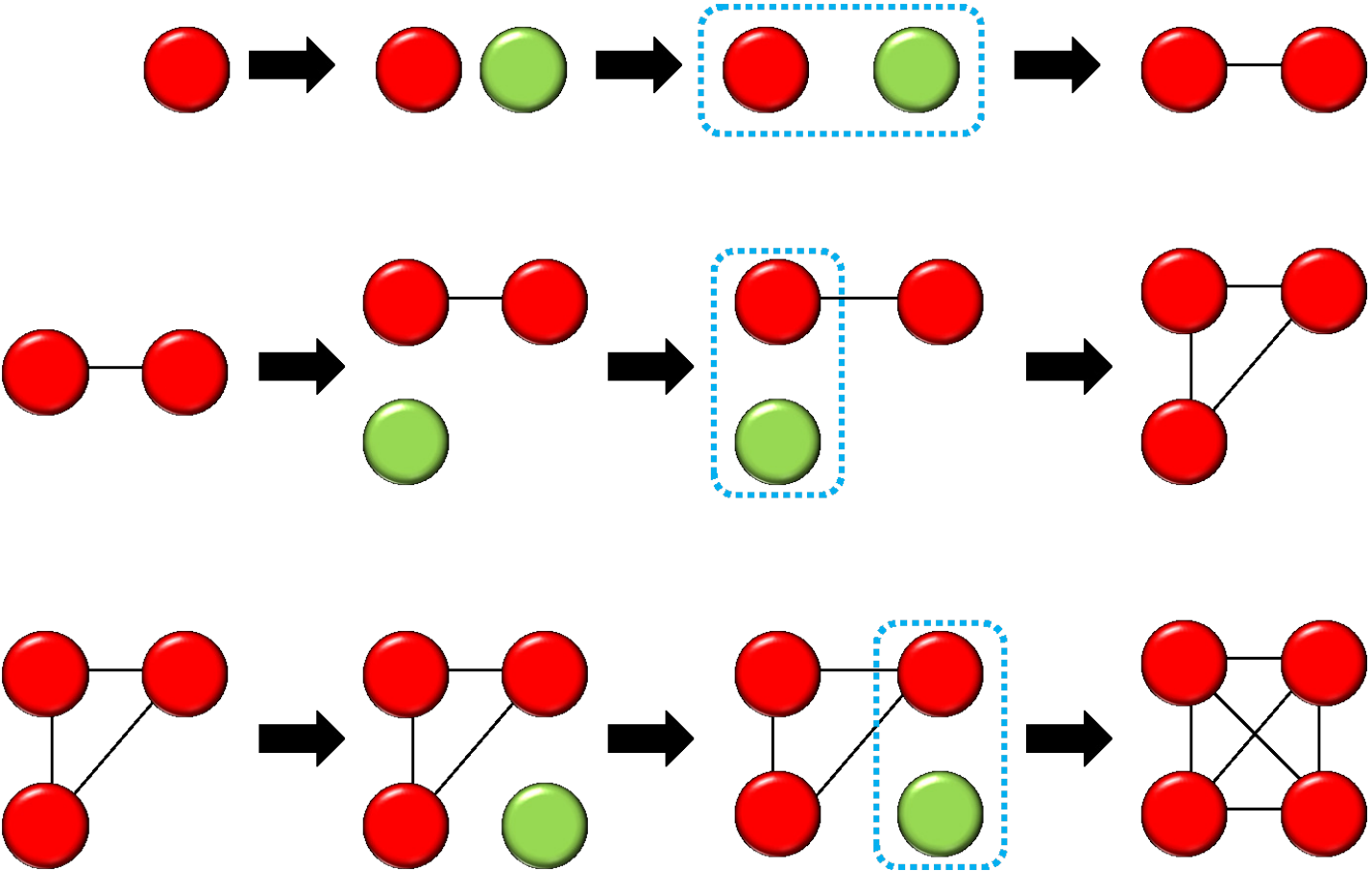}}
	\caption{Large scale $W$-state preparation strategy, starting from initially separable qubits in $|1\rangle$ and $|0\rangle$ states in the first step (first row), shown with red and green circles, respectively. Blue dashed rectangle represent the expansion process and the ancillary qubit is not shown for clarity. A 3-qubit $W$-like state is prepared in the second step (middle row) which is then expanded to a genuine $|W_4\rangle$ state in the third step (last row).}\label{fig:Concept}
\end{figure}

In Fig.~\ref{fig:Concept}, we illustrate the first three steps of our strategy for preparing a large scale $W$ state, requiring no initial entanglement.
We start with a qubit (red circle) in $|1\rangle$ state.
In the first row, applying the basic expansion operation (blue dashed rectangle) to the qubit with an additional qubit (green circle) in $|0\rangle$ state (and the ancillary qubit in $|0\rangle$ state, not shown in the figure) a $W$-type Einstein-Podolsky-Rosen (EPR) pair is created.
In the middle row, the EPR pair is expanded with another qubit in $|0\rangle$ state, creating a 3-qubit $W$-like state in the form
${1 \over \sqrt{4}}|100\rangle+{1 \over \sqrt{4}}|010\rangle + {1 \over \sqrt{2}} |001\rangle$.
The third step is to expand the 3-qubit $W$-like state with another qubit in $|0\rangle$ state, obtaining a 4-qubit $W$ state in the form
${1 \over \sqrt{4}}|1000\rangle+{1 \over \sqrt{4}}|0100\rangle + {1 \over \sqrt{4}} |0010\rangle + {1 \over \sqrt{4}} |0001\rangle$.
It is straightforward to continue expanding the $|W_4\rangle$ state in the same fashion.

Besides starting with qubits initially in a separable state, it is also possible to double the size of an existing $n$-qubit $W$ state.
Assuming that each intermediate step for expanding the state by one qubit is appropriately achieved -which can also be done in parallel, we now move to a more intuitive notation.
We consider that the initial state is in the form
$|W_n\rangle |0\rangle^{\otimes n} |0\rangle_{Anc}^{\otimes n}$,
which requires swapping qubits appropriately so that each triple of qubits shall consist of one qubit of the $W$ state, one ancillary qubit, and one qubit to join the $W$ state, respectively.
For $n=3$, the overall expansion process can be described as
\begin{equation}\label{:scale3}
\left.\begin{aligned}
&\hskip0.5cm  |W_3\rangle |0\rangle^{\otimes 3} |0\rangle_{Anc}^{\otimes 3} \\
& \to tr_{2,5,8}[ O^{\otimes 3} ( SW_{5,9} SW_{3,4} SW_{2,7})  (|W_3\rangle |0\rangle^{\otimes 3} |0\rangle_{Anc}^{\otimes 3}) ] \\
& =|W_6\rangle,
\end{aligned}\right.
\end{equation}
\noindent where $tr_i$ denotes tracing out $i$'th qubit and $SW_{i,j}$ denotes swapping $i$'th and $j$'th qubits. For $n$ qubits, denoting the overall qubit swap operations as SWAP, the expansion process reads

\begin{equation}\label{:scale4}
\left.\begin{aligned}
& \hskip0.5cm  |W_n\rangle |0\rangle^{\otimes n} |0\rangle_{Anc}^{\otimes n} \\
& \to tr_{2,5,.., 3n-1}[ O^{\otimes n} \text{SWAP} (|W_n\rangle |0\rangle^{\otimes n} |0\rangle_{Anc}^{\otimes n}) ] \\
& = |W_{2n}\rangle.
\end{aligned}\right.
\end{equation}

On the contrary to inherently probabilistic fusion or expansion operations presented in Refs.~\cite{Tashima2008,Tashima2009A,Tashima2009B,Ikuta2011,Tashima2010,Ozdemir2011,Bugu2013A,Yesilyurt2013A,Ozaydin2014A,Ming16Teleportation,Diker15,Ming16PRA,CavityRefs2,RecentW4,RecentW1}, this operation is in principally deterministic, i.e. besides inevitable experimental imperfections -which also exist in fusion operations, the resultant state is predicted with certainty, requiring no post-measurement that would possibly destroy some of the qubits and shrink the size of the target state.

\section{Physical Model}

We consider an NV-center coupled to an optical cavity.
When an incident photon is introduced to the cavity, the Langevin equations can be solved and neglecting the vacuum input field in the rotating frame, assuming weak assumption limit $\langle \sigma_z \rangle = -1$ and adiabatically eliminating the cavity mode, the reflection coefficient for the input photon pulse is found as~\cite{Munro2008PRB1,Milburn94,Munro2008PRB2}

\begin{equation}\label{eq:refcoeff}
	r(\omega_p) = {  [ i (\omega_C - \omega_p) - {\kappa \over 2}  ] [ i (\omega_0 - \omega_p) + {\gamma \over 2} ] + g^2
		\over
		[ i (\omega_C - \omega_p) + {\kappa \over 2}  ] [ i (\omega_0 - \omega_p) + {\gamma \over 2} ] + g^2 },
\end{equation}

\noindent where $\omega_p$, $\omega_C$ and $\omega_0$ are the frequency of the incident photon, frequency of the cavity field and the transition frequency of the electronic energy levels, respectively. $g$ is the coupling strength of the cavity to the NV center, $\kappa$ is the cavity decay rate and $\gamma$ is the NV center decay rate.

\noindent If the NV center is uncoupled from the cavity, the reflection coefficient for the input photon becomes

\begin{equation}\label{eq:refcoeff0}
	r_0(\omega_p) = {   i (\omega_C - \omega_p) - {\kappa \over 2}
		\over
		i (\omega_C - \omega_p) + {\kappa \over 2}
	}.
\end{equation}

\noindent The reflection coefficients can be obtained for the resonant condition $\omega_p=\omega_0=\omega_C$ as

\begin{equation}\label{eq:refcoeff1}
	r(\omega_p) = {    - \kappa \gamma + 4 g^2
		\over
		\kappa \gamma + 4 g^2
	}, \ \ \text{and} \ \  r_0(\omega_p) = -1.
\end{equation}

\begin{figure}[t]
	\centerline{\includegraphics[width=0.3\columnwidth]{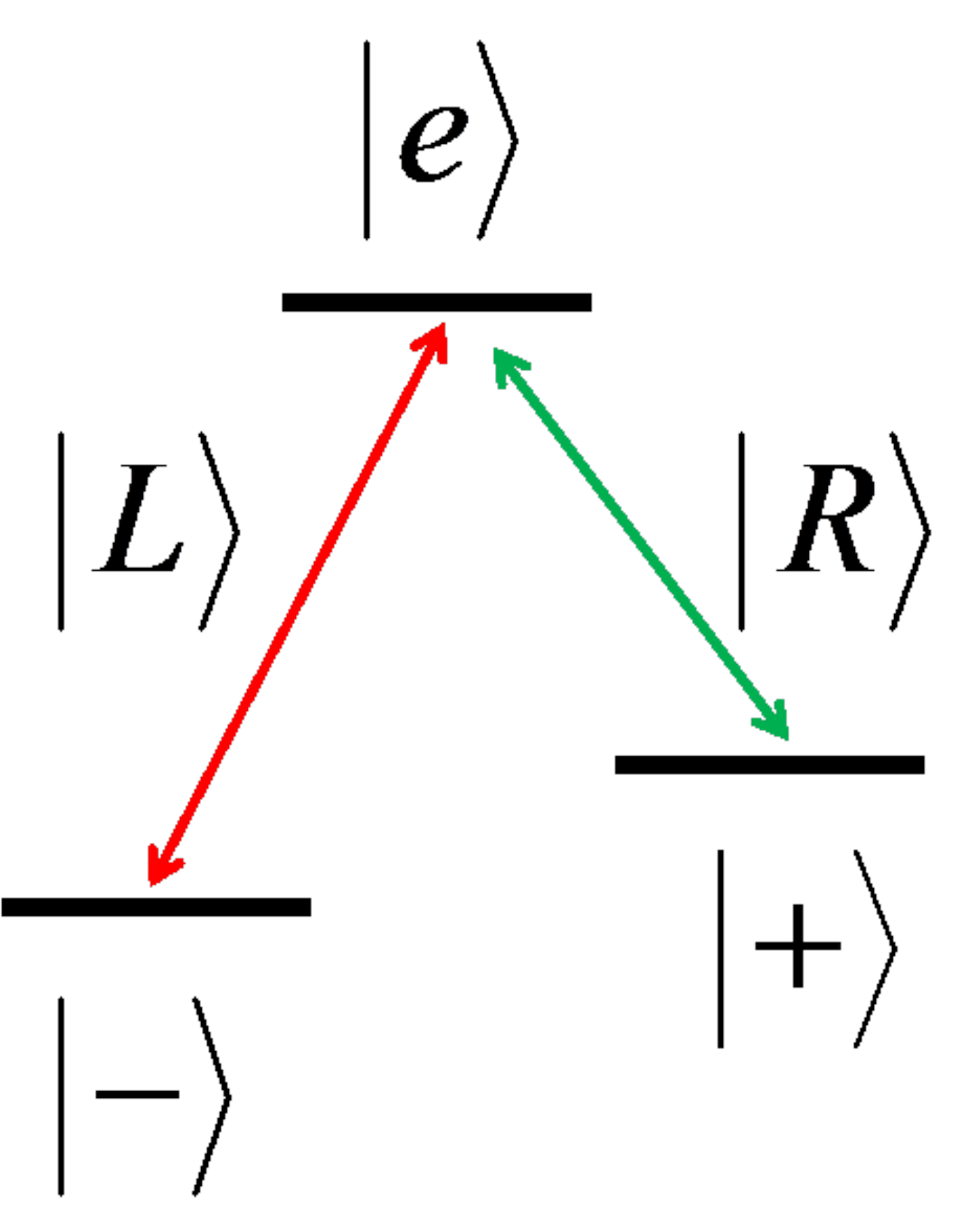}}
	\caption{$\Lambda$ type optical transitions possible in an NV center. The transitions $|-\rangle  \leftrightarrow |e\rangle$ and $|+\rangle \leftrightarrow |e\rangle$  are associated with the left and right polarization of the photon, denoted as $|L\rangle$ and $|R\rangle$ respectively.}\label{fig:excitation}
\end{figure}

$|R\rangle$ and $|L\rangle$ denoting the right and left circular polarization states, respectively; due to the spin-dependent optical transition rules~\cite{Cheng2013JOSAB} as simply illustrated in Fig.~\ref{fig:excitation} and optical Faraday rotation, an $|R\rangle$ polarized incident photon receives a phase shift $e^{i \phi_0}$ because, due to large level splitting, the spin state of the NV center is decoupled from the incident pulse~\cite{Munro2008PRB1}. However, if the incident photon is $|L\rangle$ polarized, it will receive a phase shift $e^{i \phi}$ ($e^{i \phi_0}$) depending on the spin state of the NV center $|-\rangle$ $(|+\rangle)$, where $\phi$ and $\phi_0$ are the arguments of the complex numbers $r(\omega_p)$ and $r_0(\omega_p)$, respectively. For the resonant condition, and $g > 5 \sqrt{\kappa \gamma}$, one approximately finds $\phi=0$ and $\phi_0=\pi$. Placing a $\pi$ phase shifter to the photon reflection path, a controlled-Z gate between the electronic spin of the NV center and the incident photon is realized as $|R\rangle|+\rangle \rightarrow |R\rangle|+\rangle$, $|R\rangle|-\rangle \rightarrow |R\rangle|-\rangle$, $|L\rangle|+\rangle \rightarrow |L\rangle|+\rangle$, $|L\rangle|-\rangle \rightarrow -|L\rangle|-\rangle$~\cite{Du11PRA,Cheng2013JOSAB,Yeon13OptExp}.
Note that the implementations of single qubit operations on NV center spins and incident photons are presented in the Appendix.
\begin{figure}[t]
	\centerline{\includegraphics[width=0.9\columnwidth]{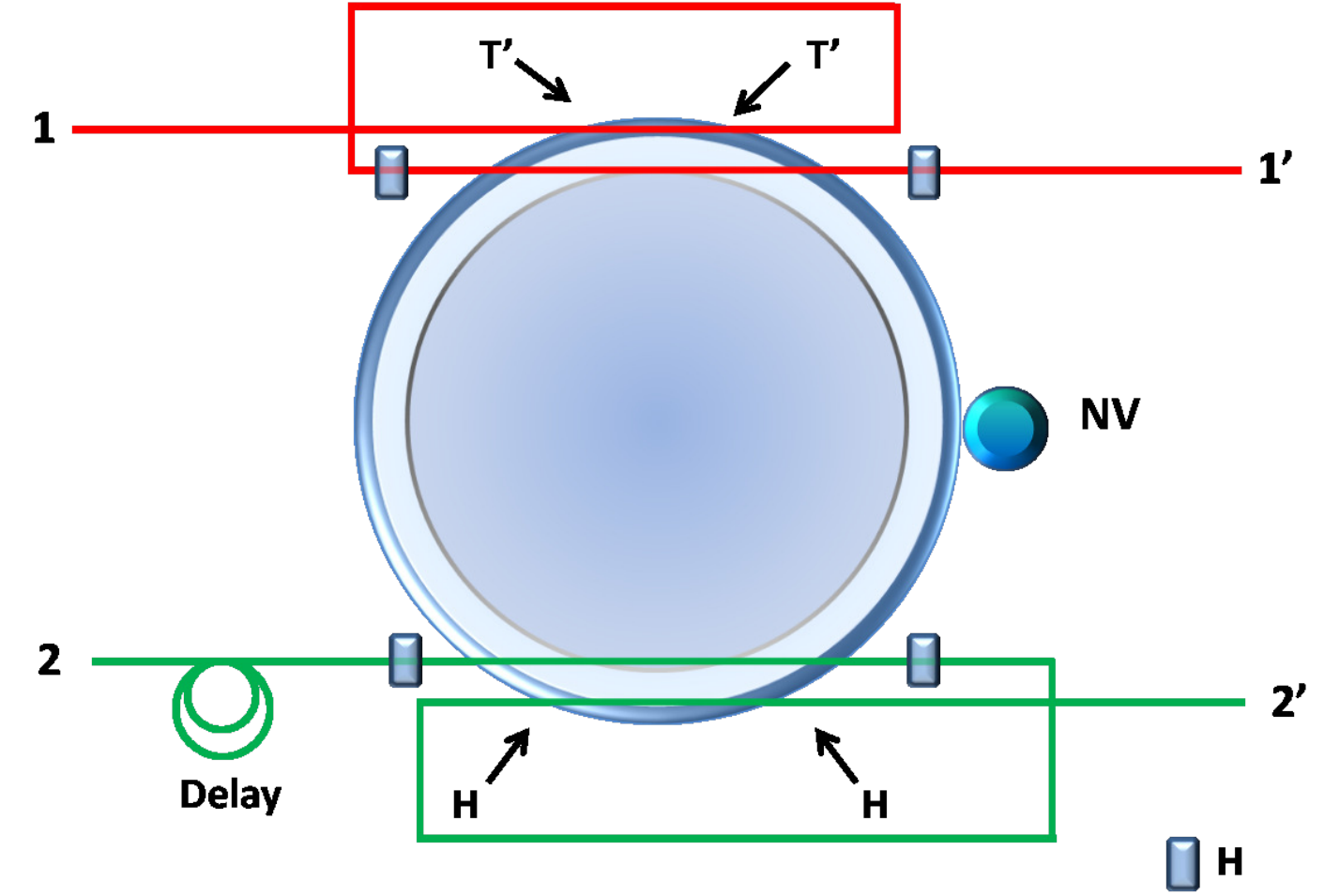}}
	\caption{NV center assisted setup for creating or expanding photonic $W$ states. $H$ denotes the Hadamard gate. Delay is to denote that the photon in input 2 arrives to the setup after an arbitrary time. The $\pi$ phase shifter on the photon output paths are not shown for clarity. For the definition $\text{T}'$ gate and overall operation of the setup, please see the text.}\label{fig:WPhotonic}
\end{figure}
\section{Creating or Expanding a Photonic $W$ State Using an NV Center}
In this section, we first present how a $W$-type photonic Bell state in the form $|W_2\rangle = {1 / \sqrt{2} } ( |L\rangle |R\rangle + |R\rangle |L\rangle)$ can be created from an initially separable state of two photons in the circular polarization states $|R\rangle$ and $|L\rangle$.
The strategy is as follows: A photon in input 1, spin in the NV center (as the ancillary qubit) and a photon in input 2 are prepared in the $|L\rangle \otimes |+\rangle \otimes |R\rangle$ state, respectively.
The setup for physical realization of the circuit model in Fig.~\ref{fig:NVabst} is presented in Fig.~\ref{fig:WPhotonic}.
A $\text{T}'_1$ gate is applied to the NV center spin, before and after the interaction between the photon in input 1.
A Hadamard gate is applied to the photon in input 1, before and after the second interaction.
Next, comes the photon in input 2.
A Hadamard gate is applied to the photon, and the NV center spin, before and after the first, and the second interactions, respectively.
In summary, starting with an initial $|L\rangle |+\rangle |R\rangle$ state and applying the gates with indices as shown in Fig.~\ref{fig:NVabst} and finally tracing out the NV spin, one obtains the state
${1 / \sqrt{2} } ( |L\rangle |R\rangle + |R\rangle |L\rangle)$.
That is, although the two photons in input 1 and input 2 have never interacted directly, a two-photon $W$-type Bell pair is created between them.\\

We now proceed with the strategy how an arbitrarily large $W$ state of $n$ photons in the form
\begin{equation}
\left.\begin{aligned}
|W_{n}\rangle  = & { 1 \over \sqrt{n} }( \> |L_1 R_2 ... R_{n-1} R_n\rangle + |R_1 L_2 ...R_{n-1} R_n\rangle \\
& \hskip0.35cm + ... +  |R_1 R_2 ... R_{n-1} L_n\rangle )
\end{aligned}\right.
\end{equation}
\noindent can be expanded to a $W$ state of $2n$ photons.
Each photon of the $W$ state is sent to the NV center in Fig.~\ref{fig:WPhotonic} together with a $|R\rangle$ polarized photon through inputs 1 and 2, respectively.
NV center spin which is initially prepared in $|+\rangle$ state is found in the same state after each operation.
Therefore, the same NV center spin can be used arbitrary times before it decoheres, or by being set back to  $|+\rangle$ state between the expansion operations.
Each operation expands the size of the state by one, creating a $W$-like state with weighted superposition coefficients, and through the final operation, a genuine $|W_{2n}\rangle$ state is obtained.

\begin{figure}[t]
	\centerline{\includegraphics[width=1\columnwidth]{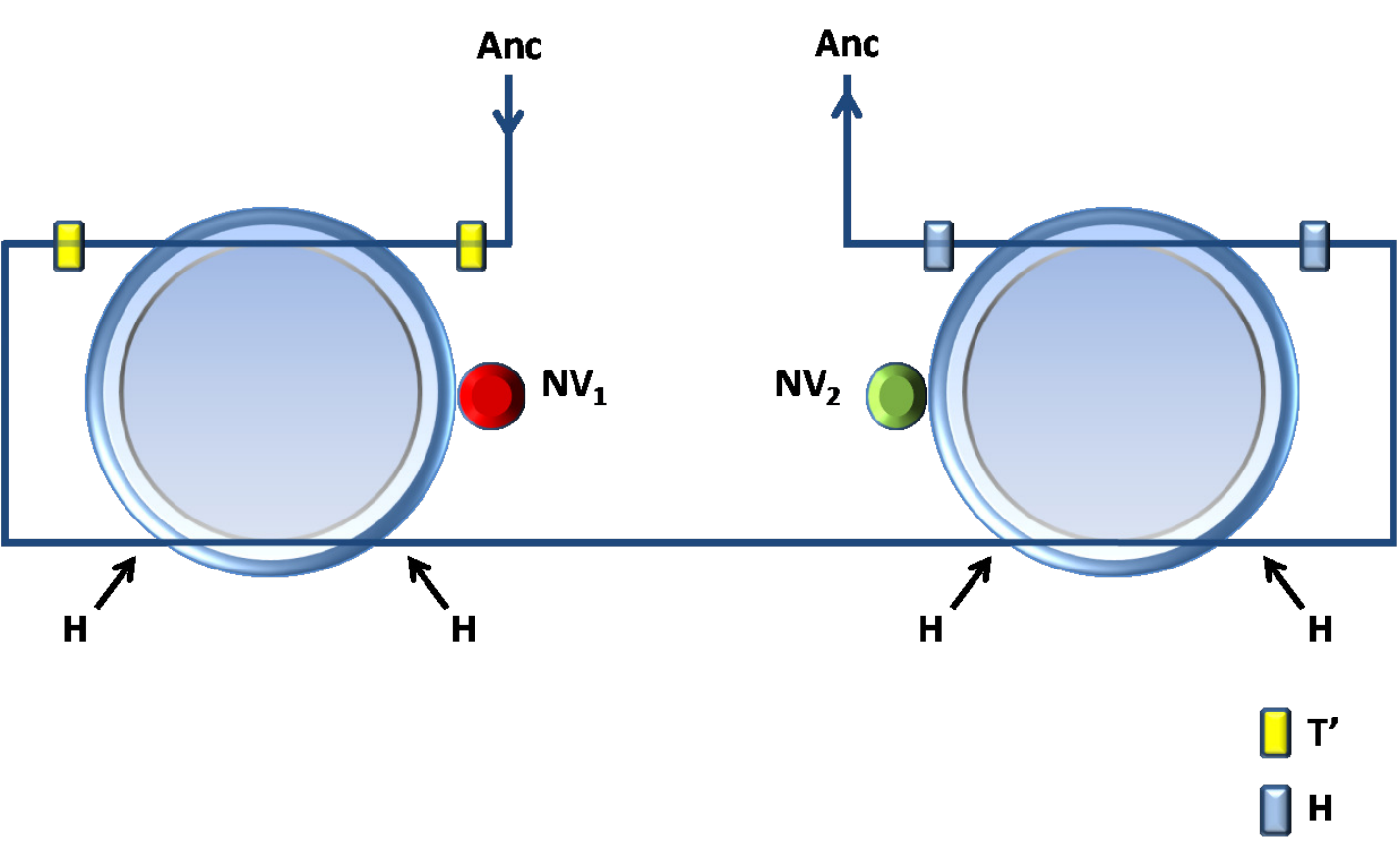}}
	\caption{Setup for realizing the basic expansion step for preparing large-scale $W$ state of NV center spins, via an ancillary photon. The photon interacts with the first NV center between two $T'$ gates defined in the text. Before and after the second interaction, Hadamard gates are applied to the NV center. Then the photon is sent to the spatially separated NV center. Before and after the first interaction, Hadamard gates are applied to the NV center and finally, Hadamard gates are applied to the photon before and after the second interaction. The $\pi$ phase shifter on the photon output paths are not shown.}
	\label{fig:WNV}
\end{figure}

\section{Creating or Expanding a $W$ State of Distant NV Center Spins Using an Ancillary Photon}
We now present how to create a two-qubit $W$-type Bell pair, and how to expand an arbitrary size $W$ state of distant NV center spins in the form
\begin{equation}
\left.\begin{aligned}
|W_{n}\rangle  = & { 1 \over \sqrt{n} }( \> |-_1 +_2 ... +_{n-1} +_n\rangle + |+_1 -_2 ... +_{n-1} +_n\rangle \\
& \hskip0.35cm + ... + |+_1 +_2 ... +_{n-1} -_n\rangle )
\end{aligned}\right.
\end{equation}
\noindent via an ancillary photon, implementing the circuit model in Fig.~\ref{fig:NVabst}.
An NV center $(NV_2)$ with the spin state $|+\rangle$ is in input 2, and the ancillary photon in the state $|R\rangle$ is in $Anc$ input.
In the $W$-type Bell pair-creation scenario, an NV center in the state $|-\rangle$, and in the $W$ state expansion scenario, an NV center spin as one of the qubits of the $W$ state is in input 1 $(NV_1)$.
As illustrated in Fig.~\ref{fig:WNV}, the photon interacts with $NV_1$ between two $T'$ gates.
Before and after the second interaction, a Hadamard gate is applied to $NV_1$.
Then the  photon is sent to $NV_2$ which is subject to Hadamard gates before and after the first interaction.
Finally, the photon interacts with $NV_2$ again, between two Hadamard gates.
It is straightforward to show that this setup realizes the circuit model in Fig.~\ref{fig:NVabst}, achieving the operation $O$ in Eq.~\ref{eq:operation}.
That is, a three-qubit system of $NV_1$, ancillary photon and $NV_2$ initially prepared in $|-\rangle|R\rangle|+\rangle$ state is transformed into
${1 / \sqrt{2}}(|-\rangle|R\rangle|+\rangle + |+\rangle|R\rangle|-\rangle)$ state, i.e. creating $W$-type Bell pair ${1 / \sqrt{2}}(|-\rangle|+\rangle + |+\rangle|-\rangle)$ between two distant NV center spins, leaving the ancillary photon in the $|R\rangle$ state ready for another operation.

\begin{figure}[t]
	\centerline{\includegraphics[width=1\columnwidth]{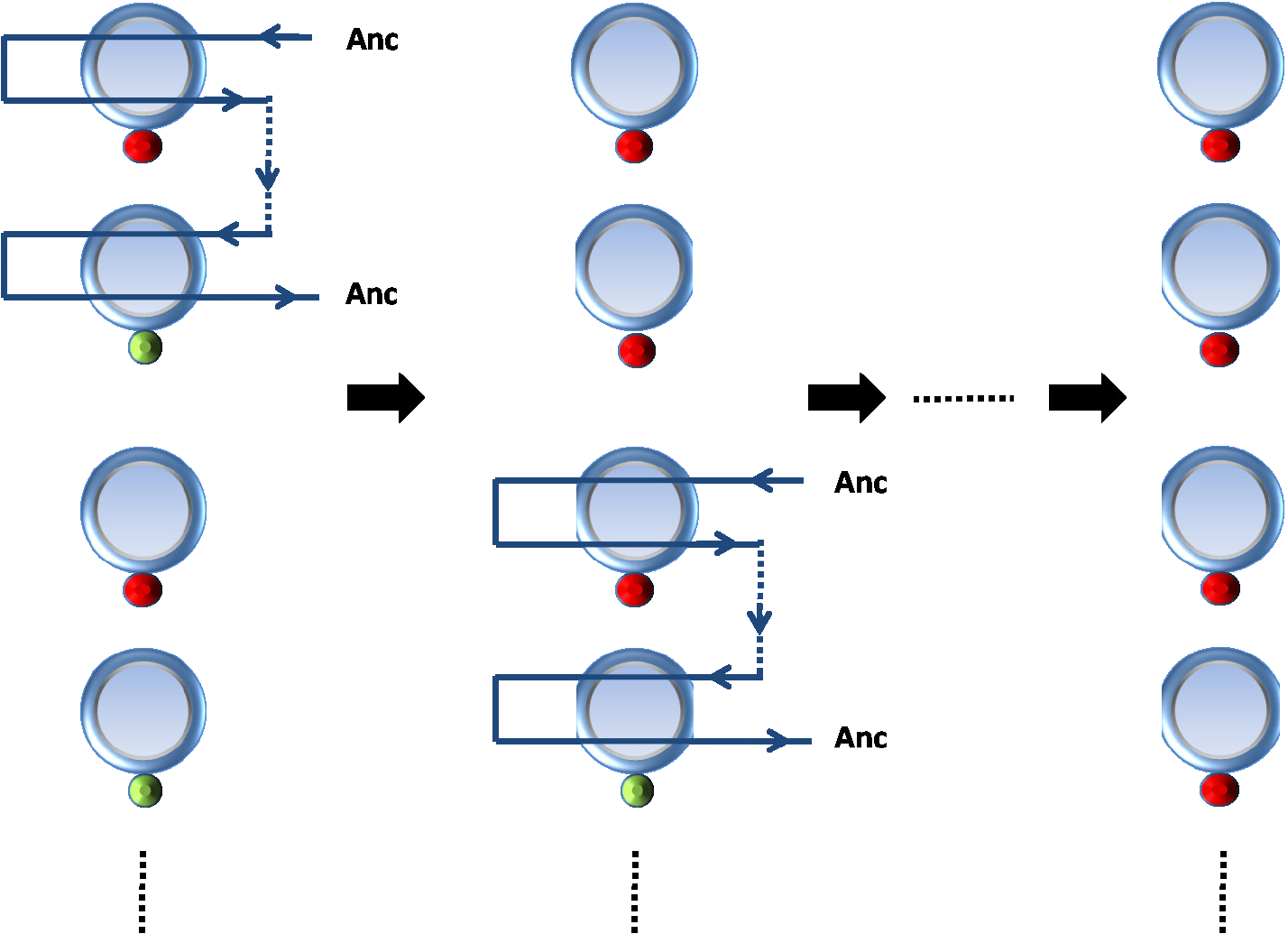}}
	\caption{An ancillary photon (Anc) in $|R\rangle$ circular polarization state is sent to a pair of NV centers, first (red circle) belonging to the $|W_n\rangle$ state to be expanded, and second (green circle) being the one to join to the $W$ state.
		Each operation expands the state by one qubit, creating $W$-like state, and with the last operation on $n$'th pair, a $|W_{2n}\rangle$ state is obtained.
		Alternatively, $n$ ancillary photons can be used to realize the expansion process in parallel.
		Single qubit gates (in Fig.~\ref{fig:WNV}) and the $\pi$ phase shifter on the photon output paths are not shown. }
	\label{fig:WNVScalable}
\end{figure}

The implementation of the strategy of Fig.~\ref{fig:Concept} for expanding a $|W_n\rangle$ state to a $|W_{2n}\rangle$ state  is as follows, and illustrated in Fig.~\ref{fig:WNVScalable}.
An ancillary photon in $|R\rangle$ state is sent to each pair of NV centers, first being a qubit of the $W$ state to be expanded, and second being the additional qubit to join the $W$ state.
Single qubit operations are applied appropriately as shown in Fig.~\ref{fig:WNV}.
Each of the $n-1$ operations expands the $W$ state by one qubit, leading to a $W$-like state with weighted superposition coefficients, as explained in Section III.
Finally, the last operation creates a genuine $|W_{2n}\rangle$ state.


Due to the flexibility of our model, the same ancillary photon can be used to apply the operations by traveling among each pair of NV centers consecutively in a serial manner,
or $n$ distinct ancillary photons can be used in parallel.
For the latter case, the transformation
$|W_{n}\rangle \otimes |R\rangle^{\otimes n} \otimes  |+\rangle^{\otimes n} \rightarrow |W_{2n}\rangle \otimes  |R\rangle^{\otimes n}$ is achieved and
a $W$ state of $2n$ NV center spins is prepared.

\section{Fidelity Analysis Due to Imperfections}
In this section, we analyze the effects of non-ideal gates of the circuit presented in Fig.\ref{fig:NVabst} on the fidelity of the prepared $W$ state.
We assume that the initial logical and ancillary qubits are prepared in the ideal state, and after the operations, the ancillary qubit is left in its ideal initial state  (or that after each round, a fresh ancillary qubit can be prepared). Because no post-measurements are required, we focus on possible imperfections in implementing controlled-Z (CZ), Hadamard and $T'$ gates.
We consider non-ideal Hadamard and $T'$ gates as follows:

\begin{equation}
\text{H} \left( \alpha \right)  = \left(
	\begin{array}{cc}
		\cos \left( \theta - \alpha\right) & \sin \left(\theta - \alpha\right) \\
		\sin \left(\theta - \alpha\right) & -\cos \left(\theta - \alpha\right) \\
	\end{array}
	\right),
\end{equation}

\begin{equation}
	\text{T}' \left( \beta \right)  = \left(
	\begin{array}{cc}
		\cos \left( \theta - \beta\right) & \sin \left(\theta - \beta\right) \\
		\sin \left(\theta - \beta\right) & -\cos \left(\theta - \beta\right) \\
	\end{array}
	\right),
\end{equation}
\noindent with $\theta = \pi/4$ and $\theta = \pi/8$. They represent ideal Hadamard and $T'$ gates for $\alpha=0$ and $\beta=0$, respectively. Similarly, a non-ideal CZ gate is considered as a general controlled-phase gate 
\begin{equation}
	\text{CP}(\gamma) = \left(
	\begin{array}{cccc}
		1 & 0 & 0 & 0 \\
		0 & 1 & 0 & 0 \\
		0 & 0 & 1 & 0 \\
		0 & 0 & 0 & \text{Exp}[i (\pi-\gamma)]
	\end{array}
	\right),
\end{equation}

\noindent representing an ideal CZ gate for $\gamma=0$. Hence, for a non-zero $\alpha$, $\beta$ or $\theta$, not a genuine $W$ state but a $W$-like state is obtained.
We calculate the fidelity of the $W$-like state obtained in the size doubling process, i.e. $|W_n\rangle \rightarrow |W_{2n}\rangle$ with respect to these imperfections. 
We first find that the fidelity does not depend on the number of qubits, $n$. 
This finding can be interpreted as follows, and be related to the robustness of $W$ states. 
Unlike a GHZ state for example, in each superposition term of a $W$ state, only one qubit is in $|1\rangle$ state. 

Therefore, although we do not measure and learn which one, each controlled-gate in the circuit in Fig.\ref{fig:NVabst} apply (ideal or non-ideal) $Z$ operation only when the logical qubit in input 1 is in $|1\rangle$, and apply an identity operator otherwise. 
In the latter case, applying two single qubit gates of the same type consecutively is equivalent to an identity operator, as well.
Hence, for an arbitrary $n$, including the case ($n=1$) where an EPR pair is created from two separable qubits, the fidelities with respect to each imperfection separately and the combined one are found as
\begin{equation*}
	F_{\text{H}}(\alpha) = \left| \frac{1}{2}  + \frac{1}{2} \cos\left(2\alpha\right) ^3 \right|^2,
\end{equation*}
\begin{equation*}
	F_{\text{T'}}(\beta) = \left| \frac{1}{\sqrt{2}}\left[ 
	\cos\left(\frac{\pi + 8 \beta}{4}  \right) 
   +\sin\left(\frac{\pi + 8 \beta}{4}  \right) 	
	\right] \right|^2,
\end{equation*}
\begin{equation*}
F_{\text{CP}}(\gamma)\!\!=\!\!\left|\frac{1}{2} e^{-2 i \gamma} \!\cos\left(\frac{\gamma}{2}\right)^4\!\! +\!\!   
\frac{1}{\sqrt{2}} \left[ \cos\left(\frac{\pi}{8}\right)^2 \!\!-\! e^{-i \gamma} \!\sin\left(\frac{\pi}{8}\right)^2 \right] \right|^2\!,
\end{equation*}	   

\begin{eqnarray*}
	\begin{split}
	 F_{\text{Combined}}&(\alpha,\beta,\gamma) =  \\
	&\left|	\frac{e^{-4 i \gamma} (1+ e^{i \gamma})^4 \cos({2\alpha})^3 \cos\left[\frac{1}{4}\left(\pi + 8 \beta\right)\right]}{16\sqrt{2}} \right.\\ 
	+&\left. \frac{e^{-i \gamma} \big\{ -1+ e^{i \gamma} + (1+ e^{i \gamma})\sin\left[ \frac{1}{4}\left(\pi + 8 \beta\right)  \right] \big\} }{2\sqrt{2}}	 \right|^2\!\!.
	\end{split}
\end{eqnarray*}

\begin{figure}[t]
	\centerline{\includegraphics[width=1\columnwidth]{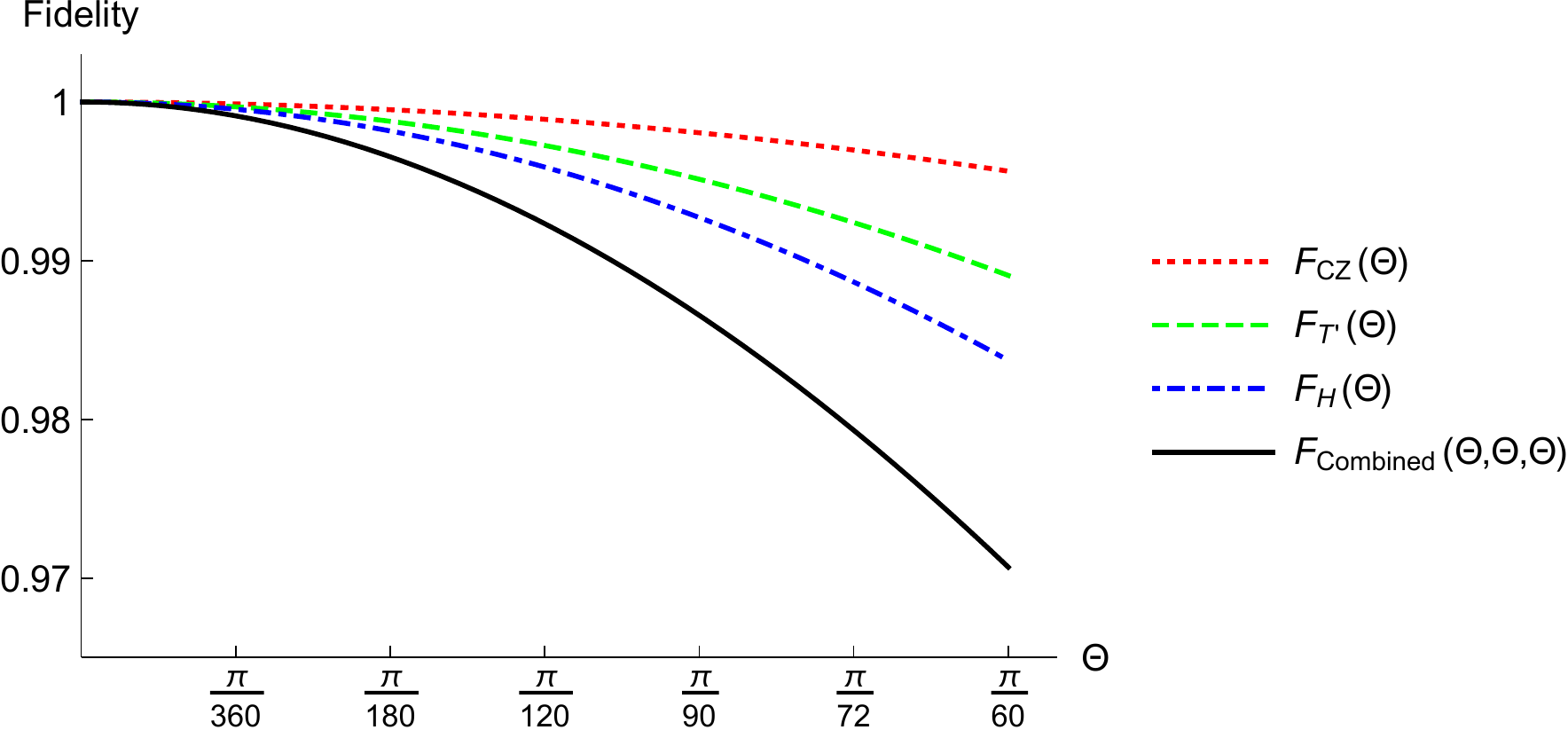}}
	\caption{Fidelity as a function of imperfection in CZ gate (dotted red curve), $T'$ gate (dashed green curve), Hadamard gate (dotdashed blue curve), and combined (solid black curve). For $\Theta=\pi/60$ with all gate the imperfections combined, fidelity is greater than $0.97$. }
	\label{fig:Fidelities}
\end{figure}

\noindent We plot the fidelity results with respect to the same parameter, $0\leq\Theta\leq \pi/60$ in Fig.\ref{fig:Fidelities}, and find that for $\Theta= \pi/60$ with all the gate imperfections combined, fidelity is greater than $0.97$.

\section{Discussion}
The analyses in Refs.\cite{Du11PRA,Long15SRep,Wang17OptExp} suggest that the present setups are feasible with the current technology.
Previous proposals and experimental demonstrations for preparing $W$ states based on fusion strategies are inherently probabilistic, requiring post-selection and possibly post-processing even in the ideal experimental conditions.
This causes a significant decrease in the efficiency of the strategies, requiring a sub-exponential resource cost in the optimal case as shown in Monte Carlo simulations in Refs.~\cite{Ozdemir2011,Ozaydin2014A}.
However, we call our schemes \emph{deterministic in principle} because in an ideal experimental realization, no matter what technology and what kind of qubits are used,  our circuit model illustrated in Fig.~\ref{fig:NVabst} operates in a deterministic way.
Besides possible experimental imperfections -which are plausibly minimized using NV centers even in the room temperature- the spin-photon interaction is not ideal due to broad phonon sideband. 
However, being deterministic in principle, the present proposal can be significantly more efficient than the proposals which are probabilistic even without including the experimental imperfections.

Another essential aspect of the present proposal is that, in most of the proposals based on spin systems, after the interaction with the incident photon, an extra photon is required to interact with spin qubit to reveal the spin state or the entangled state created between the qubits.
Our proposal on the contrary, does not require such a post-measurement or forward processing which requires additional interactions. 
Similarly to a recent work on implementing the Magic Square Game with quantum dots~\cite{BuguMSG}, the ancillary system herein is used just like a  \emph{catalyzer}, left in the initial state after the operation and can be used again.
Note that in any task based on distant spin or atomic qubits and traveling ancillary photons such as those in ~\cite{Lukin10Nature,Du11PRA,Deng13PRA,Cheng2013JOSAB,Wu13OptExp,Yeon13OptExp,Zhang13JOSAB,Long15PRA,BuguMSG},
operations on three or more qubits are implicitly realized. 
Hence, considering a three-qubit operation as the basic building block enables a more systematic approach.
The three-qubit operation $O$ presented in Eq.~\ref{eq:operation} and its applications to $W$ states of photonic or spin qubits can be used not only to expand a $|W_n\rangle$ with $n$ qubits to a $|W_{2n}\rangle$ in a deterministic way, but also to expand a $W$ state with arbitrary size of qubits in a probabilistic way with specific fidelities, following the approach presented in Ref.~\cite{DetExp}.

Note that we do not exclude the possibility of further decreasing the two-qubit operations (realized by the interaction of the incident photon with the spin).
However, such a reduction would break the in-principle-deterministic nature of our proposal, requiring post-measurements and post-processing, achievable by introducing additional two-qubit operations, resulting in the increase of the number of overall two-qubit operations.

Regarding the feasibility of our setup, experimental demonstrations of polarization independent couplers with high efficiency were reported~\cite{Bakir2020IEEE,Cheben2015OptExp}. Microtoroid resonators with whispering gallery modes (WGM) have been considered for realizing the photon-NV center spin interactions~\cite{Deng13PRA,Cheng2013JOSAB,Wu13OptExp,Yeon13OptExp,Zhang13JOSAB,Long15PRA,Wang15JOSAB,Long15SRep,Wang17OptExp}. 
Due to the non-transversality of WGM, the WGM resonators behaves differently from
conventional ring or Fabry-Perot resonators~\cite{Junge2013PRL}. 
That is, the orthogonal polarization states are correlated with the propagating directions of the photons, making the counter-propagating photons distinguishable. 
As an alternative to WGM microtoroid cavities, single-sided cavities can be utilized for realizing the interactions required in our setups.
Single- and double-sided cavities have been recently attracted attention for realizing spin-photon interactions~\cite{Wei14OptExp,Hu17SRep,Heo17SRep,Ren18AdP,MingLi18AdP}.
On the other hand, due to their relatively low Q-factors, a major issue in using single-sided optical cavities is that photon losses and effective weak measurements can decrease the fidelity of the prepared state.
However, considering the NV center in diamond in a photonic crystal (PC) modeled as a single-sided low-Q cavity, Young et al. showed that the NV center spin can be efficiently measured with high fidelity, and  that with appropriate modifications, their system can be used for entangling spatially separated NV center spins~\cite{Young2009NJP}. Achieving Q-factors up to $10^7$~\cite{Song2005NatMat}, PCs are promising as cavities for NV center spins, and experimental demonstrations have been reported~\cite{Englund2010NanoLett,Faraon2012PRL,Schukraft2016APLP,Fehler2019ACSNano}. 

An interesting future problem is to design error correction mechanisms for  systems such as the one presented herein based on spin-photon interactions.

In conclusion, we proposed a three-qubit operation for creating and expanding arbitrary size $W$ states and decomposed this operation into only two- and single-qubit gates.
We showed that this operation can be implemented for, i) photonic systems, assisted by an ancillary spin qubit, and ii) spin qubits, assisted by an ancillary photonic qubit.
We presented two exemplary setups for NV center systems in microcavities.
We analyzed the effects of experimental imperfections in implementing the gates on the fidelity of the prepared $W$ state.
\subsection*{Acknowledgments}
FO greatly acknowledges the hospitality of M. Koashi during his stay in The University of Tokyo, and the financial support from Tokyo International University Personal Research Fund. 
CY acknowledges Istanbul University Scientific Research Fund, Grant Number: BAP-2019-33825.
SB acknowledges Japanese Government MEXT scholarship. 
MK acknowledges the support by CREST (Japan Science and Technology Agency) Grant No: JPMJCR1671. \\

\section*{Appendix}
Single qubit operations on photons considered in this paper can be realized simply by an half wave plate (HWP) with the operation~\cite{AdamHWP}

\begin{equation}
\text{HWP}(\theta / 2) = \left(
      \begin{array}{cc}
        \cos\theta & \sin\theta \\
        \sin\theta & -\cos\theta \\
      \end{array}
    \right),
\end{equation}

\noindent which realizes an Hadamard gate for $\theta= \pi / 8$ and a $T'$ gate for for $\theta= \pi / 16$.

For the single qubit operations on NV center spins, EM pulses can be used as described in Refs.~\cite{Nemoto2014PRX,Nemoto2016SciRep}.
Here, we follow holonomic quantum computation which attracted much attention recently~\cite{holo1,holo2,holo3,holo4}.
In Ref.~\cite{holo4} it was demonstrated that the highest excited state $|e\rangle$ can be connected to $|-\rangle$ and $|+\rangle$ states by a single tunable laser which can generate frequency side-bands and nanosecond pulses~\cite{holo5} by electro-optical modulation.
In the rotating frame, the system is described by the Hamiltonian

\begin{equation}\label{eq:HolonomicHamiltonian}
H(t) = {\hbar \Omega(t) \over 2} ( u |e\rangle\langle-| + v |e\rangle\langle+| + h.c.) + \Delta |e\rangle\langle e|.
\end{equation}

\noindent Here, $\Omega(t)$ is the pulse envelope for both tones and the transitions associated with $|-\rangle  \leftrightarrow |e\rangle$ and $|+\rangle \leftrightarrow |e\rangle$ are scaled by complex constants $u=\sin(\theta / 2)$ and $v = - e^{- i \phi} \cos(\theta / 2) $, respectively, controlled by the tuning of the relative strength and phase between the carrier and sideband frequencies. $\Delta$ denotes the photon detuning.
The dark state of the Hamiltonian decoupled from the dynamics is $|d\rangle = \cos(\theta / 2) |-\rangle$ and the bright state which undergoes an excitation to $|e\rangle$ is $|b\rangle = \sin(\theta / 2) |-\rangle - e^{i \phi} \cos(\theta / 2) |+\rangle $. For the pulse duration $\tau = 2 \pi \sqrt{\Omega^2 + \Delta ^2}$, a purely geometric operator is realized via the cyclic, non-adiabatic evolution in $\{ |b\rangle, |d\rangle \}$ subspace as $U(\theta,\phi,\Delta / \Omega) = |d\rangle \langle d| + e^{i \gamma} |b\rangle \langle b| $, with $\gamma = \pi (1 - \Delta / \sqrt{\Omega^2 + \Delta^2})$, which performs
\begin{equation}
U(\theta) = \left(
      \begin{array}{cc}
        \cos\theta & \sin\theta \\
        \sin\theta & -\cos\theta \\
      \end{array}
    \right).
\end{equation}
\noindent for $\gamma = \pi$ and $\phi=0$ up to a global phase~\cite{holo2}.
Hence, Hadamard and $T'$ gates can be realized for $\theta = \pi / 4$ and $\theta = \pi / 8$, respectively.


\end{document}